\documentclass[prd,preprint,showpacs]{revtex4}
\usepackage{graphicx}

\usepackage{amssymb}

\def\be{\begin{equation}}
\def\ee{\end{equation}}
\def\bea{\begin{eqnarray}}
\def\eea{\end{eqnarray}}

\def\d{\rm{d}}

\def\vr{\varrho}

\newcommand{\bi} {\begin{itemize}}
\newcommand{\ei} {\end{itemize}}
\newcommand{\ben} {\begin{enumerate}}
\newcommand{\een} {\end{enumerate}}

\newcommand{\Om} {{\Omega}}

\newcommand{\Ga} {{\Gamma}}
\newcommand{\ga} {{\gamma}}

\def\v1{\vspace{1cm}}

\begin{document}
\title{Energy, angular momentum, superenergy and angular supermomentum
in conformal frames}

\author{Mariusz P. D\c{a}browski}
\email{mpdabfz@wmf.univ.szczecin.pl}
\author{Janusz Garecki}
\email{garecki@wmf.univ.szczecin.pl}
\affiliation{\it Institute of Physics, University of Szczecin, Wielkopolska 15,
          70-451 Szczecin, Poland}

\date{\today}

\begin{abstract}
We find the rules of the conformal transformation
for the energetic quantities such as the Einstein energy-momentum
complex, the Bergmann-Thomson angular momentum complex, the
superenergy tensor, and the angular supermomentum tensor of gravitation and matter.

We show that the conformal transformation rules for the matter
parts of both the Einstein complex and the Bergmann-Thomson complex are
fairly simple, while the transformation rules for their gravitational parts
are more complicated. We also find that the transformational rules of
the superenergy tensor of matter and the superenergy tensor of gravity are
quite complicated except for the case of a pure gravity. In such a special case
the superenergy density as well as the sum of the superenergy density and the matter
energy density are invariants of the conformal transformation. Besides, in that case,
a conformal invariant is also the Bel-Robinson tensor which is a part of the
superenergy tensor. As for the angular supermomentum tensor of gravity - it emerges that
its transformational rule even for a pure gravity is quite complicated
but this is not the case for the angular supermomentum tensor of
matter.

Having investigated some technicalities of the conformal transformations, we also
find the conformal transformation rule for the
curvature invariants and, in particular, for the Gauss-Bonnet invariant in
a spacetime of arbitrary dimension.
\end{abstract}

\pacs{98.80.-k;98.80.Jk;04.20.Cv;04.50.-h}

\maketitle


\section{Introduction}
\label{sect1}

\setcounter{equation}{0}

It is widely known that there exists a problem of the energy-momentum
of gravitational field in general relativity. Since
the gravitational field may locally vanish, then one is
always able to find the frame in which the energy-momentum of the gravitational
field vanishes in this frame, while it may not vanish in the other frames. In fact,
the physical objects which describe this situation are not tensors,
and they are called the gravitational field pseudotensors. They
form the energy-momentum complexes which are the sums of the obvious
energy-momentum tensors of matter and appropriate pseudotensors.
The choice of the gravitational field pseudotensor
is not unique so that many different definitions of the
pseudotensors have been proposed. To our knowledge, the most frequently used are
the energy-momentum complexes of Einstein \cite{einstein},
Landau-Lifshitz \cite{landau},
M\o ller \cite{Mol72}, Papapetrou \cite{papapetrou}, Bergmann-Thomson \cite{BT},
Weinberg \cite{Weinberg}, and Bak-Cangemi-Jackiw \cite{Jackiw}.
Among them only Landau-lifshitz, Weinberg,
and Bak-Cangemi-Jackiw pseudotensors are symmetric.
On the other hand, only the Einstein complex is canonical.
Bearing in mind the formalism of field theory, it is obvious that there
also exist the angular momentum complexes. Among
them the Bergmann-Thomson angular
momentum complex \cite{BT} and the Landau-Lifshitz angular momentum complex \cite{landau}
are most widely used.

The arbitrariness of the choice of pseudotensors and the
fact that they usually give different results for the same
spacetime inspired some authors \cite{superenergy,supermomentum,supertensors}
to define quantities which describe the generalized energy-momentum
content of the gravitational field in a tensorial way.
These quantities are called gravitational
superenergy tensors \cite{superenergy} and gravitational angular supermomentum tensors
\cite{supermomentum} or super$^{(k)}$-energy tensors
\cite{supertensors}.

The canonical superenergy tensors and the canonical
angular supermomentum tensors have successfully been calculated
for plane, plane-fronted and cylindrical gravitational waves,
Friedmann universes, Schwarzschild, and Kerr spacetimes
\cite{superenergy,supermomentum}.

By use of the superenergy and
angular supermomentum tensors one can also prove that a real gravitational wave
with $R_{iklm}\not= 0$ possesses and carries positive-definite
superenergy and angular supermomentum \cite{superenergy,supermomentum,cqg05}.

In Refs. \cite{paper1,paper2}, the properties of the Einstein and
Bergmann-Thomson complexes as well as the superenergy and
supermomentum tensors for G\"odel spacetime were studied. It was
shown that the former were sensitive to a particular choice of
coordinates while the latter were coordinate-independent. Some
interesting properties of superenergy and supermomentum were
found. For example, the relation between the positivity of the
superenergy and causality violation.

In this paper we are going to discuss another interesting problem
- the sensitivity of the complexes, superenergy and supermomentum
to the conformal transformations of the metric tensor
\cite{hawk_ellis}. Conformal transformations are interesting
characteristics of the scalar-tensor theories of gravity
\cite{bd,maeda,scaltens,polarski}, including its conformally
invariant version \cite{jordan,flanagan,faraoni,annalen07}. The point is
that these theories can be represented in the two conformally
related frames: the Jordan frame in which the scalar field is
non-minimally coupled to the metric tensor, and in the Einstein
frame in which it is minimally coupled to the metric tensor. It is
most striking that the scalar-tensor theories of gravity are the
low-energy limits of the currently considered as the most fundamental unification
of interactions theory such as superstring theory
\cite{polchinsky,superjim,veneziano,quevedo}. It has been shown
that some physical processes such as the universe inflation
and density perturbations look different in conformally related frames
\cite{veneziano,quevedo}. This is the main motivation why we find interesting to investigate
the problem of energy-momentum in the context of conformal
transformations. Inspired by similar ideas, in Ref. \cite{tekin} it
was shown that Arnowitt-Deser-Misner (ADM) masses were invariant in different conformal
frames for asymptotically AdS and flat spacetimes as long as the
conformal factor goes to unity at infinity.

Our paper is organized as follows. In Section II we give basic
review of the idea of the conformal transformations of the metric
tensor and discuss the transformational properties of the
geometric quantities such as, for example, the Gauss-Bonnet
invariant in $D-$spacetime dimensions. In Section III we discuss
the conformal transformation of the Einstein energy-momentum
complex while in the Section IV the conformal transformation of
the Bergmann-Thomson angular momentum complex. In Section V we
discuss the conformal transformation of the superenergy tensors
and in Section VI the conformal transformations of the angular
supermomentum tensors. In Section VII we give our conclusions.

\section{Basic properties of conformal transformations}
\label{conrel}
\setcounter{equation}{0}

Consider a spacetime $({\cal M}, g_{ab})$, where ${\cal M}$ is a smooth
$n-$dimensional manifold and $g_{ab}$ is a Lorentzian metric on $M$.
The following conformal transformation
\bea
\label{conf_trafo}
\tilde{g}_{ab}(x) &=& \Omega^2(x) g_{ab}(x)~,
\eea
where $\Omega$ is a smooth, non-vanishing function of the spacetime point
is a point-dependent rescaling of the metric and is called a conformal factor. It must
be a twice-differentiable function of coordinates $x^{k}$ and
lie in the range $0<\Omega<\infty$ ($a,b,k,l = 0,1,2, \ldots D$).
The conformal transformations shrink or stretch the distances between the two points
described by the same coordinate system $x^{a}$ on the manifold
${\cal M}$, but they preserve the angles between vectors
(in particular null vectors which define light cones) which
leads to a conservation of the (global) causal structure of the manifold \cite{hawk_ellis}.
If we take $\Om =$ const. we deal with the so-called scale
transformations \cite{maeda}. In fact, conformal transformations are
localized scale transformations $\Om = \Om(x)$.

On the other hand, the coordinate transformations $x^{a} \to \tilde{x}^{a}$ only
change coordinates and do not change geometry so that
they are entirely different from conformal transformations \cite{hawk_ellis}.
This is crucial since conformal
transformations lead to a different physics \cite{maeda}.
Since this is usually related to a different coupling
of a physical field to gravity, we will be talking about different
frames in which the physics is studied (see also Refs. \cite{flanagan,faraoni} for
a slightly different view).

In $D$ spacetime dimensions the
determinant of the metric $g={\rm det}~g_{ab}$ transforms as
\bea
\label{det}
\sqrt{-\tilde{g}} &=& \Omega^D \sqrt{-g}~.
\eea
It is obvious from (\ref{conf_trafo}) that the following relations
for the inverse metrics and the spacetime intervals hold
\bea
\label{conf_trafo_inv}
\tilde{g}^{ab} &=& \Omega^{-2} g^{ab}~, \\
\d \tilde{s}^2 &=& \Omega^2 \d s^2~.
\eea
Finally, the notion of conformal flatness means that
\bea
\label{conf_flat}
\tilde{g}_{ab} \Omega^{-2}(x) &=& \eta_{ab}~,
\eea
where $\eta_{ab}$ is the flat Minkowski metric.

The application of (\ref{conf_trafo}) to the Christoffel connection coefficients
gives \cite{hawk_ellis}
\bea
\label{connections}
\tilde{\Ga}^{c}_{ab}
&=&
\Ga^{c}_{ab} + \frac{1}{\Om}\left( g^{c}_{a} \Om_{,b} +
  g^{c}_{b} \Om_{,a} - g_{ab}g^{cd}\Om_{,d} \right)~,
  \hspace{0.5cm}\tilde{\Ga}^{b}_{ab}
= \Ga^{b}_{ab} + D \frac{\Om_{,a}}{\Om}
\\
\label{connections1}
\Ga^{c}_{ab}
&=&
\tilde{\Ga}^{c}_{ab}- \frac{1}{\Om} \left( \tilde{g}^{c}_{a}
  \Om_{,b} + \tilde{g}^{c}_{b} \Om_{,a} -
  \tilde{g}_{ab}\tilde{g}^{cd}\Om_{,d} \right)~,
  \hspace{0.5cm}\tilde{\Ga}^{b}_{ab}
= \Ga^{b}_{ab} - D \frac{\Om_{,a}}{\Om}.
\eea

The Riemann tensors, Ricci tensors, and Ricci scalars in the two related frames $g_{ab}$ and
$\tilde{g}_{ab}$ transform as \footnote{We use the sign convention (-+...++), the Riemann
tensor convention $R^a_{~bcd} = \Gamma^a_{bd,c} - \Gamma^a_{bc,d} +
\Gamma^a_{ce} \Gamma^e_{bd} - \Gamma^a_{de} \Gamma^e_{cb}$, and the Ricci tensor is $R_{bd}
= R^{a}_{~bad}$.}
\bea
\label{riemanntensor1}
\tilde{R}^a_{~bcd} &=& R^a_{~bcd} + \frac{1}{\Om}\left[\delta^a_d
\Om_{;bc} - \delta^a_c \Om_{;bd} + g_{bc} \Om^{;a}_{~;d} - g_{bd}
\Om^{;a}_{~;c} \right] \\
&+& \frac{2}{\Om^{2}} \left[\delta^a_c
\Om_{,b} \Om_{,d} - \delta^a_d \Om_{,b} \Om_{,c} + g_{bd} \Om^{,a}
\Om_{,c} - g_{bc} \Om^{,a} \Om_{,d} \right]
+ \frac{1}{\Om^{2}} \left[\delta^a_d g_{bc} - \delta^a_c g_{bd}
\right]g_{ef} \Om^{,e} \Om^{,f} ~~, \nonumber
\\
\label{riemanntensor2}
R^a_{~bcd} &=& \tilde{R}^a_{~bcd} - \frac{1}{\Om}\left[\delta^a_d
\Om_{\tilde{;}bc} - \delta^a_c \Om_{\tilde{;}bd} + \tilde{g}_{bc} \Om^{\tilde{;}a}_{~\tilde{;}d}
- \tilde{g}_{bd} \Om^{\tilde{;}a}_{~\tilde{;}c} \right] \\
&+& \frac{1}{\Om^{2}} \left[\delta^a_d \tilde{g}_{bc} - \delta^a_c \tilde{g}_{bd}
\right]\tilde{g}_{ef} \Om^{,e} \Om^{,f} ~~, \nonumber
\eea
\bea
\label{riccitensor1}
\tilde{R}_{ab}
&=&
R_{ab} + \frac{1}{\Om^{2}}\left [
  2(D-2)\Om_{,a}\Om_{,b}-(D-3) \Om_{,c}\Om^{,c}g_{ab}\right ]
-\frac{1}{\Om}\left [ (D-2)\Om_{;ab}+ g_{ab} \Box \Om  \right ]~,
\\
\label{riccitensor2}
R_{ab} &=& \tilde{R}_{ab} - \frac{1}{\Om^2}(D-1)\tilde{g}_{ab} \Om_{,c}\Om^{,c}
+\frac{1}{\Om}\left[(D-2)\Om_{\tilde{;}ab}+ \tilde{g}_{ab} \stackrel{\sim}{\Box} \Om  \right ]~,
\eea
\bea
\label{ricciscalar4}
\tilde{R} &=& \Om^{-2} \left [ R - 2(D-1)\frac{\Box{\Om}}{\Om} -
(D-1)(D-4) g^{ab} \frac{\Om_{,a}\Om_{,b}}{\Om^2}
\right]~,\\
\label{ricciscalar5}
R &=& \Omega^2 \left[ \tilde{R} + 2(D-1)
\frac{\stackrel{\sim}{\Box}\Om}{\Om} - D(D-1) \tilde{g}^{ab}
\frac{\Om_{,a}\Om_{,b}}{\Om^2}\right ]~,
\eea
and the appropriate d'Alambertian operators change under (\ref{conf_trafo}) as
\bea
\stackrel{\sim}{\Box}\phi &=&\Om^{-2}\left( {\Box}\phi+
  (D-2)g^{ab}\frac{\Om_{,a}}{\Om}\phi_{,b} \right )~,\\
\Box\phi &=&\Om^{2}\left( \stackrel{\sim}{\Box}\phi-
  (D-2)\tilde{g}^{ab}\frac{\Om_{,a}}{\Om}\phi_{,b} \right )~.
\eea
In these formulas the d'Alembertian $\stackrel{\sim}{\Box}$ taken with respect to the
metric $\tilde{g}_{ab}$ is different
from $\Box$ which is taken with respect to a conformally rescaled metric
$g_{ab}$. Same refers to the covariant derivatives $\tilde{;}$
and $;$ in (\ref{riemanntensor1})-(\ref{riccitensor2}).

For the Einstein tensor we have
\bea
\label{Eintensor1}
\tilde{G}_{ab}
&=&
G_{ab} + \frac{D-2}{2\Om^2}\left [
  4\Om_{,a}\Om_{,b}+(D-5) \Om_{,c}\Om^{,c}g_{ab}\right ]
-\frac{D-2}{\Om} \left [ \Om_{;ab} - g_{ab} \Box \Om  \right ]~,
\\
\label{Eintensor2}
G_{ab} &=& \tilde{G}_{ab} - \frac{D-2}{2\Om^2}(D-1)\Om_{,e}\Om^{,e} \tilde{g}_{ab}
+ \frac{D-2}{\Om} \left[\Om_{\tilde{;}ab} - \tilde{g}_{ab} \stackrel{\sim}{\Box}\Om  \right]~,
\eea

An important feature of the conformal transformations is that they
preserve Weyl conformal curvature tensor $(D \geq 3)$
\bea
\label{weyl_def}
C_{abcd} &=& R_{abcd} +
\frac{2}{D-2}\left(g_{a[d}R_{c]b} +
g_{b[c}R_{d]a}\right) + \frac{2}{(D-1)(D-2)} R
g_{a[c}g_{d]b}~,
\eea
which means that we have (note that one index is raised)
\bea
\label{weyl_trafo}
\tilde{C}^{a}_{~bcd} &=& C^{a}_{~bcd}
\eea
under (\ref{conf_trafo}).
Using this property (\ref{weyl_trafo}) and the rules (\ref{conf_trafo})-(\ref{conf_trafo_inv})
one can easily conclude that the Weyl Lagrangian \cite{weyl}
\be
\tilde{L}_w = - \alpha \sqrt{-\tilde{g}} \tilde{C}^{abcd} \tilde{C}_{abcd}
 = - \alpha \sqrt{-g} C^{abcd} C_{abcd} = L_w~
\ee
is an invariant of the conformal transformation
(\ref{conf_trafo}).

In further considerations it would be useful to know the conformal transformation rules
for the widely applied curvature invariants which are given by
\bea
\label{R2}
\tilde{R}^2 &=& \Om^{-4} \left[ R^2 + 4 (D-1)^2 \Om^{-2} \left( \Box \Om
\right)^2 + (D-1)^2 (D-4)^2 \Om^{-4} g^{ab} \Om_{,a} \Om_{,b}
g^{cd} \Om_{,c} \Om_{,d} \right. \nonumber \\ &-& \left.
4 (D-1) R \Om^{-1} \Box \Om - 2 R
(D-1)(D-4) \Om^{-2} g^{ab} \Om_{,a} \Om_{,b} \right. \nonumber \\
&+& \left. 4(D-1)^2(D-4)
\Om^{-3} \Box \Om g^{ab} \Om_{,a} \Om_{,b} \right]~,
\eea
\bea
\label{RabRab}
\tilde{R}_{ab}\tilde{R}^{ab} &=& \Om^{-4} \left\{ R_{ab}R^{ab} - 2
\Om^{-1} \left[(D-2)R_{ab} \Omega^{;ab} + R \Box \Om \right] \right. \nonumber \\
&+& \left. \Om^{-2} \left[ 4(D-2) R_{ab} \Om^{,a} \Om^{,b} -
2(D-3) R \Om_{,e} \Om^{,e} + (D-2)^2 \Om_{;ab} \Om^{;ab} + (3D -
4) \left( \Box \Om \right)^2 \right] \right. \nonumber \\
&-& \left. \left[ (D-2)^2 \Om_{;ab} \Om^{,a} \Om^{,b} -
(D^2 - 5D + 5) \Box \Om \Om_{,e} \Om^{,e} \right] \right.
\nonumber \\ &+& \left. \Om^{-4} (D-1) (D^2 - 5D + 8) \left(\Om_{,a} \Om^{,a} \right)^2
\right\}~,
\eea
\bea
\label{Riem2}
\tilde{R}_{abcd} \tilde{R}^{abcd} &=& \Om^{-4} \left\{R_{abcd} R^{abcd}
- 8 \Om^{-1} R_{bc} \Omega^{;bc} + 4 \Om^{-2} \left[\left( \Box \Om \right)^2
+ (D-2)\Om_{;bc} \Om^{;bc} - R \Om_{,b} \Om^{,b} \right. \right.
\nonumber \\ &+& \left. \left. 4 R_{bc} \Om^{,b} \Om^{,c}
\right] + 8 \Om^{-3} \left[(D-3) \Box \Om \Om_{,c} \Om^{,c}
- 2(D-2)\Om_{;bc} \Om^{,b} \Om^{,c} \right] \right. \nonumber \\
&+& \left. 2 \Om^{-4} D(D-1) \left(\Om_{,a} \Om^{,a} \right)^2  \right\}~.
\eea

In fact out of these curvature invariants one forms the well-known Gauss-Bonnet
term which is one of the Euler (or Lovelock) densities \cite{lovelock,GB}. Its
conformal transformation (\ref{conf_trafo}) reads as
\bea
\label{RGBtil}
\tilde{R}_{GB} &\equiv& \tilde{R}_{abcd} \tilde{R}^{abcd} - 4
\tilde{R}_{ab} \tilde{R}^{ab} + \tilde{R}^2
= \Omega^{-4} \left\{ R_{GB} + 4(D-3) \Omega^{-1} \left[ 2
R_{ab} \Omega^{;ab} - R \Box \Om \right] \right. \nonumber \\
&+& \left. 2 (D-3) \Om^{-2} \left[2(D-2) \left( \left( \Box \Om \right)^2
- \Om_{;ab} \Om^{;ab} \right) - 8 R_{ab} \Om^{,a} \Om^{,b} -
(D-6) R \Om_{,a} \Om^{,a} \right] \right. \nonumber \\
&+& \left. 4(D-2)(D-3) \Om^{-3} \left[(D-5) \Box \Om \Om_{,a} \Om^{,a}
+ 4 \Om_{;ab} \Om^{,a} \Om^{,b} \right] \right. \nonumber \\
&+& \left. (D-1)(D-2)(D-3)(D-8) \Om^{-4} \left(\Om_{,a} \Om^{,a} \right)^2
\right\}~.
\eea
The inverse transformation is given by
\bea
\label{RGB}
R_{GB} &\equiv& R_{abcd} R^{abcd} - 4 R_{ab} R^{ab} + R^2 =
\Om^4 \left\{ \tilde{R}_{GB} - 4(D-3) \Om^{-1}
\left[2\tilde{R}_{ab} \Om^{\tilde{;} ab} - \tilde{R} \stackrel{\sim}{\Box} \Om
\right] \right. \nonumber \\
&+& \left. 2(D-2)(D-3) \Om^{-2} \left[ 2\left( \stackrel{\sim}{\Box} \Om
\right)^2 - 2\Om_{\tilde{;}ab} \Om^{\tilde{;}ab} - \tilde{R} \Om_{\tilde{;}a}
\Om^{\tilde{;}a} \right] \right. \\
&-& \left. (D-1)(D-2)(D-3) \Om^{-3} \left[4 \left( \stackrel{\sim}{\Box} \Om
\right) \Om_{\tilde{;}a} \Om^{\tilde{;}a} - D \Om^{-1}
\left( \Om_{\tilde{;}a} \Om^{\tilde{;}a} \right)^2 \right] \right\}~.
\nonumber
\eea

So far we have considered only geometrical part. For the
matter part we usually consider matter lagrangian of the form
\bea
\tilde{S}_m = \int{ \sqrt{-\tilde{g}} \Omega^{-D} d^D x {\cal{L}}_{\rm m} }=
\int{ \sqrt{-g} d^D x {\cal{L}}_{\rm m} }= S~,
\eea
where the conformal transformation (\ref{conf_trafo}) have been
used \cite{superjim}. Then, the energy-momentum tensor of matter
in one conformal frame reads as
\bea
\label{mattertrafotilde}
\tilde{T}^{ab} &=&
\frac{2}{\sqrt{-\tilde{g}}} \frac{\delta}{\delta \tilde{g}_{ab}}
\left(\sqrt{-\tilde{g}} \Om^{-D} {\cal{L}}_{\rm m} \right) =
  \Omega^{-D} \frac{2}{\sqrt{-g}} \frac{\partial g_{cd}}{\partial
  \tilde{g}_{ab}} \frac{\delta}{\delta g_{cd}} \left(\sqrt{-g}
  {\cal{L}}_{\rm m} \right)~,
\eea
which under (\ref{conf_trafo}) transforms as
\bea
\label{tensorlaw1}
\tilde{T}^{ab} &=& \Omega^{-D-2} T^{ab}~,\hspace{0.8cm}
\tilde{T}^a_{~b} = \Omega^{-D} T^a_{~b}\\
\label{tensorlaw2}
\tilde{T} &=& \Om^{-D} T~.
\eea
For the matter in the form of the perfect fluid with the
four-velocity $v^{a}$ ($v_{a}v^{b}=-1$), the energy density $\vr$
and the pressure ${\rm p}$
\bea
\label{enmom}
T^{ab} &=& (\vr+{\rm p})v^{a}v^{b}+{\rm p}g^{ab}~,
\eea
the conformal transformation gives
\bea
\label{enmom_trafo}
\tilde{T}^{ab}
&=&
(\tilde{\vr}+\tilde{{\rm p}})\tilde{v}^{a}\tilde{v}^{b} +
\tilde{{\rm p}}\tilde{g}^{ab}~,
\eea
where
\bea
\label{mattertrafo}
T^{ab} &=& \frac{2}{\sqrt{-g}}\frac{\delta}{\delta  g_{ab}}\left(
  \sqrt{-g}{\cal{L}}_{\rm m}\right )~,
\eea
and
\bea
\tilde{v}^{a} &=& \frac{\d x^{a}}{\d \tilde{s}} =
\frac{1}{\Om}\frac{\d x^{a}}{\d s}=\Om^{-1}v^{a}~.
\eea
Therefore, the relation between the pressure and the energy density in the conformally
related frames reads as
\bea
\tilde{\vr} &=& \Om^{-D} \vr~,\\
\tilde{{\rm p}} &=& \Om^{-D} {\rm p}~.
\eea
It is easy to note that the imposition of the conservation law in the first frame
\bea
\label{laws}
T^{ab}_{~~;b}=0~,
\eea
gives in the conformally related frame
\bea
\label{tildelaws}
 \tilde{T}^{ab}_{~~\tilde{;}b} &=& -\frac{\Om^{,a}}{\Om}\tilde{T}~.
\eea
From (\ref{tildelaws}) it appears obvious that the conformally
transformed energy-momentum tensor is conserved only, if the trace of
it vanishes ($\tilde{T}=0$) \cite{jordan,Weinberg,maeda,annalen07}. For example,
in the case of barotropic fluid with
\bea
\label{barotropic}
{\rm p}&=&(\ga-1)\vr~ \hspace{0.5cm} \gamma = {\rm const.},
\eea
it is conserved only for the radiation-type fluid
${\rm p}=[1/(D-1)]\vr$~.

Similar considerations are also true if we first impose the
conservation law in the second frame
\bea
\label{laws1}
\tilde{T}^{ab}_{~~\tilde{;}b}=0~,
\eea
which gives in the conformally related frame (no tildes)
\bea
\label{tildelaws1}
T^{ab}_{~~;b} &=& \frac{\Om^{,a}}{\Om}T~.
\eea
Finally, it follows from (\ref{tensorlaw2}) that that vanishing of
the trace of the energy-momentum tensor in one frame necessarily
requires its vanishing in the second frame, i.e., if $T=0$ in one frame, then
$\tilde{T}=0$ in the second frame and vice versa. This means only the traceless type
of matter fulfills the requirement of energy conservation.

\section{Einstein energy-momentum complex in conformal frames}
\label{sect3}

\setcounter{equation}{0}

Following our earlier work \cite{paper1,paper2} we consider
the canonical double-index energy-momentum complex
\cite{einstein,landau,Mol72}
\be
\label{Ecom1}
_E K_i^{~k} = \sqrt{- g}\bigl(T_i^{~k} + _Et_i^{~k}\bigr),
\label{complex}
\ee
where $T^{ik}$ is a symmetric energy-momentum
tensor of matter which appears on the right-hand side of the Einstein
equations and $g$ is the determinant of the metric
tensor. In fact, its definition results from the rewritten
Einstein equations \cite{Mol72}
\begin{equation}
\label{Ecom2}
\sqrt{- g}\bigl(T_i^{~k} + _E t_i^{~k}\bigr) = {_F
U_i^{~kl}}_{,l},
\end{equation}
where
\begin{equation}
_F U_i^{~kl} = - _F U_i^{~lk} = \alpha{g_{ia}\over\sqrt{-
g}}\bigl[(-g)\bigl(g^{ka}g^{lb} - g^{la}g^{kb}\bigr)\bigr]_{,b}
\label{freud}
\end{equation}
are Freud's superpotentials and
\begin{eqnarray}
_E t_i^{~k} &=& \alpha\Bigl\{\delta^k_i
g^{ms}\bigl(\Gamma^l_{mr}\Gamma^r_{sl} -
\Gamma^r_{ms}\Gamma^l_{rl}\bigr)\nonumber \\
&+& g^{ms}_{~~,i}\bigl[\Gamma^k_{ms} - {1\over 2}\bigl(\Gamma^k_{tp}
g^{tp} - \Gamma^l_{tl} g^{kt}\bigr)g_{ms} - {1\over
2}\bigl(\delta^k_s\Gamma^l_{ml} +
\delta^k_m\Gamma^l_{sl}\bigr)\bigr]\Bigr\}
\label{einpseudo}
\eea
is the Einstein's gravitational energy-momentum pseudotensor
($\alpha = c^4 / 16\pi G$).
It is useful to apply the property of metricity
\be
g^{ms}_{~~;i} = g^{ms}_{~~,i} - \Gamma^m_{ia} g^{as} - \Gamma^s_{ia} g^{ma} = 0~,
\ee
to (\ref{einpseudo}) and reduce it to a simpler form, i.e.,
\begin{eqnarray}
_E t_i^{~k} &=& \alpha\Bigl\{\delta^k_i
g^{ms}\bigl(\Gamma^l_{mr}\Gamma^r_{sl} -
\Gamma^r_{ms}\Gamma^l_{rl}\bigr)
+ \Gamma^a_{ia} \bigl( \Gamma^k_{tp} g^{tp} - \Gamma^l_{tl} g^{kt} \bigr)
\nonumber \\
&+& \Gamma^l_{ml} \bigl( \Gamma^m_{ia} g^{ak} + \Gamma^k_{ia} g^{am} \bigr)
- \Gamma^k_{ms} \bigl( \Gamma^m_{ia} g^{as} + \Gamma^s_{ia} g^{am} \bigr)
\Bigr\}.
\label{einpseudo1}
\eea
Now, we apply the conformal transformation (\ref{conf_trafo})
to the Einstein pseudotensor (\ref{einpseudo1}). We assume that
we start with the Einstein pseudotensor in the conformal frame
(with tildes) and express it in the conformal frame (no tildes) as
below
\bea
\label{pseudoE}
_E \tilde{t}_i^{~k} &=& \Omega^{-2} \Bigl\{ _E t^k_i
+ \Omega^{-1}(D-2)\bigl[ \delta_i^k \bigl(\Gamma^l_{rl}\Omega^{,r}
- g^{ms} \Gamma^r_{ms} \Omega_{,r} \bigr) + \Omega_{,i}
\bigl(g^{tp} \Gamma^k_{tp} - g^{kt} \Gamma^l_{tl} \bigr)\\
&+& \bigl( g^{ak} \Gamma^m_{ia} \Omega_{,m} + \Gamma^{k}_{ia}
\Omega^{,a} \bigr) - 2 \Gamma^{a}_{ia} \Omega^{,k} \bigr] +
\Omega^{-2}(D-1)(D-2) \bigl[\delta_i^k \Omega_{,r}\Omega^{,r} -
2 \Omega_{,i} \Omega^{,k} \bigr]  \Bigr\}~~. \nonumber
\eea
Note that in $D=2$ the rule of transformation is very simple:
$_E \tilde{t}_i^{~k} = \Omega^{-2} _E t^k_i$, which seem to reflect
the fact of the conformal flatness of all the two-dimensional
manifolds. On the other hand, it is not so simple in the case of a flat space
where all the Christoffel connection coefficients vanish.
In a general case it is clear that the transformed Einstein pseudotensor would
differ from the starting one. This seems to be natural
conclusion in the context of its sensitivity to a change of
coordinates (this is why we call it "pseudotensor") despite we do not
change those coordinates here at all (cf. Section \ref{conrel}).

From (\ref{det}) and (\ref{pseudoE}) it follows that
\bea
\label{pseudoEdet}
 \sqrt{-\tilde{g}} _E \tilde{t}_i^{~k} &=& \Omega^D \sqrt{-g} _E \tilde{t}_i^{~k}
\\
&=& \Omega^{D-2} \sqrt{-g} \Bigl\{ _E t^k_i
+ \Omega^{-1}(D-2)\bigl[ \delta_i^k \bigl(\Gamma^l_{rl}\Omega^{,r}
- g^{ms} \Gamma^r_{ms} \Omega_{,r} \bigr) + \Omega_{,i}
\bigl(g^{tp} \Gamma^k_{tp} - g^{kt} \Gamma^l_{tl} \bigr) \nonumber \\
&+& \bigl( g^{ak} \Gamma^m_{ia} \Omega_{,m} + \Gamma^{k}_{ia}
\Omega^{,a} \bigr) - 2 \Gamma^{a}_{ia} \Omega^{,k} \bigr] +
\Omega^{-2}(D-1)(D-2) \bigl[\delta_i^k \Omega_{,r}\Omega^{,r} -
2 \Omega_{,i} \Omega^{,k} \bigr]  \Bigr\}~~ \nonumber
\eea
under conformal transformation (\ref{conf_trafo}).
On the other hand, the transformation rule of the material part of
the Einstein complex (\ref{Ecom2}) reads as (cf. (\ref{det}) and
(\ref{tensorlaw1}))
\be
\label{mTik}
\sqrt{-\tilde{g}} \tilde{T}_i^k = \sqrt{-g} T_i^k~,
\ee
i.e., the quantity $\sqrt{-g} T_i^k$ is an invariant of the
conformal transformation (\ref{conf_trafo}).
Besides, if the initial metric is
Minkowskian, then the transformational rule (\ref{pseudoE}) simplifies
since the terms which contain $\Gamma^i_{kl}$ vanish.

\section{Bergmann-Thomson Angular Momentum Complex in Conformal Frames}
\label{sect4}

The Bergmann-Thomson angular momentum complex
\be
\label{BTdef}
_{BT}  M^{ijk} = -{} _{BT} M^{jki} = -{} _{BT} M^{jik}
\ee
consists of the sum of the material part
\begin{equation}
_m  M^{ijk} = \sqrt{-g}\bigl(x^i{}T^{jk} -
x^j{}T^{ik}\bigr),
\end{equation}
and the gravitational part
\begin{eqnarray}
\label{BTdeffull}
_g M^{ijk}&=& \sqrt{- g}\bigl(x^i{}_{BT} t^{jk} -
x^j{}_{BT} t^{ik}\bigr)\nonumber\\
&+& {\alpha\over\sqrt{- g}}\biggl[(-g)\bigl(g^{kj}g^{il}
- g^{ki} g^{jl}\bigr)\biggr]_{,l},
\end{eqnarray}
where
\begin{equation}
\sqrt{-g}_{BT} t^{jk} := \sqrt{-g} g^{ji}{} _E
t_i^{~k} + g^{ij}_{~~,l}{} _F U_i^{~[kl]}
\end{equation}
is the Bergmann-Thomson energy-momentum
pseudotensor of the gravitational field \cite{BT} (see also \cite{garecki01}).

As a consequence of the local energy-momentum conservation law
\begin{equation}
_E K_i^{~k}{}_{,k} = 0,
\end{equation}
which immediately follows from (\ref{Ecom2}), the angular momentum
complex satisfies local conservation laws
\begin{equation}
_{BT} M^{ijk}{}_{,k} = 0.
\end{equation}
Under the conformal transformation (\ref{conf_trafo}), the material part $_m M^{ijk}$ and
the gravitational part $_g M^{ijk}$ of the complex transform as
\begin{equation}
\label{mMijk}
_m\tilde{M}^{ijk} = \Omega^{-2}{} _m M^{ijk},
\end{equation}
\begin{eqnarray}
\label{gMijk}
_g \tilde{M}^{ijk} &=& _g M^{ijk}
+\Omega^{-1}\biggl\{2\alpha\sqrt{-g}\bigl(x^i g^{jl} -
x^j g^{il}\bigr)P_l^{~k}\nonumber\\
&+& \Omega_{,b}\bigl[4\alpha\bigl(x^i g^{lj}_{~~,t} -x^j
g^{li}_{~~,t}\bigr)g_{la}{} U^{[kt]ab} - 2\bigl(x^i{}_F
U^{j[kb]}\nonumber\\
&-& x^j{}_F U^{i[kb]}\bigr) + 4\alpha U^{[ji]kb}\bigr]\biggr\} +
\alpha\Omega^{-2}\biggl\{6\sqrt{-g}\bigl(x^i
g^{jl}\nonumber\\
&-& x^j g^{il}\bigr) Q_l^{~k} - 8
\Omega_{,b}\Omega_{,t}\bigl[(-g)\biggl(\bigl(x^i g^{kj} - x^j
g^{ki}\bigr)g^{tb}\nonumber\\
&-& \bigl(x^i g^{tj} - x^j
g^{ti}\bigr)g^{kb}\biggr)\bigr]\biggr\},
\end{eqnarray}
where
\begin{eqnarray}
P_i^{~k}&:=& \delta^k_i\bigl(\Gamma^l_{~rl}{}g^{rd}\Omega_{,d} -
g^{ms}{}\Gamma^l_{~ms}\Omega_{,l}\bigr) +
\Omega_{,i}\bigl(\Gamma^k_{~tp}{}g^{tp}\nonumber\\
&-& \Gamma^l_{~tl}{}g^{kt}\bigr) -
2\Gamma^m_{~im}{}g^{kp}\Omega_{,p} +
\Gamma^k_{~ia}{}g^{as}\Omega_{,s} +
\Gamma^m_{~ia}{}g^{ak}\Omega_{,m},
\end{eqnarray}
and
\begin{eqnarray}
Q_i^{~k}&:=&\delta_i^k{}g^{ms}\Omega_{,m}\Omega{,s} -
2g^{kp}\Omega_{,i}\Omega_{,p},\nonumber\\
U^{[kt]ab}&:=& {1\over\sqrt{-g}}\bigl[(-g)\bigl(g^{ka}
k^{tb} - g^{ta} g^{kb}\bigr)\bigr],\nonumber\\
_F  {U_i^{~[km]}}&=& {\alpha\over\sqrt{-g}}g_{ia}\bigl[(-g)\bigl(g^{ka} g^{mb} - g^{ma}
g^{kb}\bigr)\bigr]_{,b},\nonumber\\
_F U^{i[kb]}& :=& g^{il} _F{U_l^{~[kb]}}.
\end{eqnarray}
As we can see from above formulas the conformal transformation rule (\ref{mMijk}) for
the matter part of the Bergmann-Thomson angular momentum complex is fairly
simple. Let us also mention that the transformation (\ref{mMijk}) holds
in any dimension of spacetime, while the transformation (\ref{gMijk})
holds only in $D=4$ dimensions. Besides, if the initial metric is
Minkowskian, then the transformational rule (\ref{gMijk}) further simplifies
since the terms which contain $\Gamma^i_{kl}$, $g_{ik,l}$
and $g^{ik}_{~~,l}$ vanish. In this case also $_g{}M^{ijk} = 0$.

\section{Superenergy tensors in conformal frames}
\label{sect5}

\setcounter{equation}{0}

Following \cite{superenergy} one is able to introduce the canonical superenergy tensor.
The definition of the superenergy tensor $S_a^{~b}(P)$, which can be applied to an arbitrary
gravitational as well as matter field is \cite{paper1}
\begin{equation}
\label{Sabtet}
S_{(a)}^{~~~(b)}(P) = S_a^{~b}(P) := \displaystyle\lim_{\Omega\to
P}{\int\limits_{\Omega}\biggl[T_{(a)}^{~~~(b)}(y) -
T_{(a)}^{~~~(b)}(P)\biggr]d\Omega\over
1/2\int\limits_{\Omega}\sigma(P;y) d\Omega},
\label{averaging}
\end{equation}
where
\begin{eqnarray}
T_{(a)}^{~~~(b)}(y) &:= &
T_i^{~k}(y)e^i_{(a)}(y)e^{(b)}_k(y),\nonumber\\
\label{tetrad}
T_{(a)}^{~~~(b)}(P) &:=& T_i^{~k}(P)e^i_{(a)}(P)e^{(b)}_k(P) = T_a^{~b}(P)
\end{eqnarray}
are the tetrad components of a tensor or a
pseudotensor field $T_i^{~k}(y)$ which describe an energy-momentum, $y$
is the collection of normal coordinates {\bf NC(P)} at a given point
{\bf P}, $\sigma(P,y)$ is the world-function, $e^i_{(a)}(y),~~e^{(b)}_k(y)$ denote an orthonormal tetrad
field and its dual, respectively, $e^i_{(a)}(P) =
\delta^i_a,~~e^{(a)}_k(P) = \delta^a_k$, $e^i_{(a)}(y)e^{(b)}_i(y) =
\delta^b_a$, and they are paralell propagated along
geodesics through {\bf P}. At {\bf P} the tetrad and normal
components of an object are equal. We apply this
and omit tetrad brackets for indices of
any quantity attached to the point {\bf P}; for example, we write
$T^{ab}(P)$ instead of $T^{(a)(b)}(P)$ and so on.

Firstly, the symmetric superenergy tensor of matter is given by
(from now on we will also interchangeably
use $\nabla$ to mark the covariant differentiation)
\cite{paper1}
\begin{equation}
_m S_a^b(P) = \delta^{lm}\nabla_{(l}\nabla_{m)} T_a^b.
\label{SabTab}
\end{equation}
In terms of the four-velocity of an observer taken as comoving
$v^l = \delta^l_0$ $(v^lv_l = -1)$, a more convenient covariant form of (\ref{SabTab}) is
\begin{equation}
\label{Sabmatter}
_m S_a^{~ b}(P;v^l) = h^{lm}
\nabla_{(l}\nabla_{m)}{} T_a^{~ b}~,
\end{equation}
where
\be
\label{hlm}
h^{lm} \equiv 2v^lv^m + g^{lm} = h^{ml}~.
\ee

Secondly, the canonical superenergy tensor of the gravitational field
reads as
\begin{equation}
\label{Sabgrav}
_g S_a^{~b}(P;v^l) = h^{lm} {{W}_a^{~b}}{}_{lm},
\end{equation}
where
\begin{eqnarray}
\label{Tablm}
{{W}_a^{~b}}{}_{lm} & = & {2\alpha\over 9}\bigl[{B}^b_{~alm} +
{ P}^b_{~alm} \nonumber \\
& - & {1\over 2}\delta_a^b {R}^{ijk}_{~~~m}{}\bigl({R}_{ijkl} +
{R}_{ikjl}\bigr) + 2\delta_a^b
{R}_{(l\vert g}{} {R}^g_{~\vert m)}\nonumber \\
 & - & 3 {R}_{a(l\vert}{} {R}^b_{~\vert m)}
+ 2 {R}^b_{~(ag)(l\vert}{} { R}^g_{~\vert m)}\bigr],
\end{eqnarray}
and
\begin{equation}
\label{bel}
B^b_{~alm} := 2R^{bik}_{~~~(l\vert}{} R_{aik\vert m)} - {1\over 2}\delta^b_a{}
R^{ijk}_{~~~l}{} R_{ijkm} ,
\end{equation}
is the {\it Bel--Robinson tensor}, while
\begin{equation}
\label{pel}
P^b_{~alm} := 2R^{bik}_{~~~(l\vert }{} R_{aki\vert m)} - {1\over 2}\delta_a^b
{} R^{ijk}_{~~~l}{} R_{ikjm}.
\end{equation}
In vacuum, the gravitational superenergy tensor (\ref{Sabgrav}) reduces to a simpler
form:
\begin{equation}
_g S_a^{~b}(P;v^l) = {8\alpha\over 9} h^{lm} \bigl[{C^{b(ik)}}_{(l\vert}C_{aik\vert m)}
-\frac{1}{2} \delta_a^b{C^{i(kp)}}_{(l\vert}C_{ikp\vert m)}\bigr]~.
\end{equation}
It is symmetric and the quadratic form $_g S_{ab}(P;v^l)v^av^b$ is
positive-definite.

It is suggested \cite{superenergy,paper1} that the superenergy tensor $_g S_a^{~b}(P;v^l)$
should be taken as a quantity which may serve as the energy-momentum tensor for
the gravitational field. Its advantage is that it is a conserved quantity in vacuum.
The disadvantage is that the superenergy tensors
$_g S_a^{~b}(P;v^l)$ and $_m S_a^{~b}(P;v^l)$ have the
dimension: [the dimension of the components of an
energy-momentum tensor (or pseudotensor)] $\times m^{-2}$. This means it is rather that
their flux gives the appropriate energy-momentum tensors or pseudotensors.
However, in some other approach one is able to introduce average relative energy and
angular momentum tensors which have proper dimension
\cite{garecki07}. In fact, these new tensors differ from the
superenergy and angular supermomentum tensors by a constant
dimensional factor of (length)$^2$.

Now, we consider the conformal transformations of the superenergy
tensors. For the matter superenergy tensor one has
\bea
\label{mSab}
_m \tilde{S}_i^k &=& \tilde{h}^{lm}
 \tilde{\nabla}_{(l}\tilde{\nabla}_{m)}{} \tilde{T}_i^{~k} \\
&=& \Omega^{-6} {_m}S_i^k + \Omega^{-2} h^{lm} \left[2
\left( \Omega^{-4}\right)_{;(l\vert} T^{~k}_{i~~ \vert ; m )} -
( \Omega^{-4})_{;ml} T_i^k \right] \nonumber \\
&+& \Omega^{-3}
h^{lm} \left[2 {\cal D}^k_{~t(l \vert} T^{~t}_{i~~ \vert}
(\Omega^{-4})_{;m)} - {\cal D}^{p}_{~lm} T_i^k (\Omega^{-4} )_{;p}
- {\cal D}^{p}_{~li} T_p^k (\Omega^{-4} )_{;m} \right] \nonumber \\
&+& \Omega^{-6} h^{lm} \left[(P^k_{~tm} T_i^t )_{,l} - (P^t_{~mi} T_t^k )_{,l}
\right] + \Omega^{-7} h^{lm} \left[\Gamma^k_{lp} ({\cal
D}^{p}_{~tm} T_i^{t} - {\cal D}^t_{~mi} T_t^{p} )
\right. \nonumber \\ &-& \left.
\Gamma^p_{lm} ({\cal
D}^{k}_{~tp} T_i^{t} - {\cal D}^t_{~ip} T_t^{k} )
- \Gamma^p_{li} ({\cal
D}^{k}_{~mt} T_p^{t} - {\cal D}^t_{~mp} T_t^{k} )
- {\cal D}^p_{~lm} T_{i~;p}^k + {\cal D}^k_{~lp} T_{i~;m}^{p}
- {\cal D}^p_{~li} T_{p~;m}^k
\right] \nonumber \\
&+& \Omega^{-8} h^{lm} \left[{\cal D}^k_{~lp} ({\cal D}^{p}_{~tm}
T_i^t - {\cal D}^t_{~mi} T_t^p )
- {\cal D}^p_{~lm} ({\cal D}^{k}_{~tp}
T_i^t - {\cal D}^t_{~ip} T_t^k )
- {\cal D}^p_{~li} ({\cal D}^{k}_{~mt}
T_p^t - {\cal D}^t_{~mp} T_t^k )
\right]~, \nonumber
\eea
where
\be
\label{Pabc}
P^a_{~bc} = P^a_{~cb} \equiv \Omega^{-1} {\cal D}^{a}_{~bc}~,
\ee
and
\be
\label{Dabc}
{\cal D}^{a}_{~bc} \equiv \delta_b^a \Omega_{,c} + \delta_c^a
\Omega_{,b} - g_{bc} g^{ad} \Omega_{,d}~.
\ee
For the gravitational superenergy tensor one has
\begin{eqnarray}
\label{gSab}
_g\tilde{S}_a^{~b} &=& \Omega^{-4} {_g}S_a^{~b}+{2\alpha\over
9}h^{lm}\biggl\{\bigl[\bigl(g^{k[b}{}\Omega^{i]}_{~~l}-\delta^{[b}_{~~l}{}\Omega^{i]k}\bigr)
+\bigl(g^{i[b}{}\Omega^{k]}_{~~l}-\delta^{[b}_{~~l}{}\Omega^{k]i}\bigr)\bigr]
g_{[a[k}{}\Omega_{i]m]}\nonumber\\
&-&{\delta^b_a\over 2}\bigl(g^{k[i}{}\Omega^{j]}_{~~l}
-\delta^{[i}_{~~l}{}\Omega^{j]k}\bigr)\bigl(g_{[i[k}{}\Omega_{j]m]}
 + g_{[i[j}{}\Omega_{k]m]}\bigr)\biggr\}\nonumber\\
&+& {\alpha\over
9}\Omega^{-1}h^{lm}(\Omega^{-1})_{;mc}\bigl(g^{bc}\Omega_{al}
+\delta_a^b\Omega^c_{~l} -2\delta^b_l \Omega_a^{~c} -2\delta_a^c
\Omega^b_{~l}\nonumber\\
&+& g_{la} \Omega^{bc} +\delta^c_l{}\Omega^b_{~a}\bigr) +
{4\alpha\over 9}\Omega^{-2} h^{lm}\bigl[ 2R^{b(ik)}_{~~~~~l}
g_{[a[k}{}\Omega_{i]m]}\nonumber\\
&+&\bigl(g^{k[b}{}\Omega^{i]}_{~~l}-\delta^{[b}_{~~l}{}\Omega^{i]k}\bigr)R_{a(ik)m}-{\delta_a^b\over
2}\bigl(g^{k[i}{}\Omega^{j]}_{~~~l}-\delta^{[i}_{~~l}{}\Omega^{j]k}\bigr)R_{i(kj)m}\nonumber\\
&+&4\delta^b_a(\Omega^{-1})_{;lg}{}(\Omega{-1})_{;mc}g^{gc} +
6(\Omega^{-1})_{;al}(\Omega^{-1})_{;m}^{~~b}\nonumber\\
&+&{1\over 8}\bigl(R^b_{~m}{}\Omega_{al} +
\delta_a^b{}R^g_{~m}{}\Omega_{gl}
-2\delta^b_l{}R^g_{~m}{}\Omega_{ag}
-2R_{am}\Omega^b_{~l}
+ g_{la}{}R^g_{~m}\Omega^b_{~g} +
R_{lm}\Omega^b_{~a}\bigr)\bigr]\nonumber\\
&+&{4\alpha\over
3}\Omega^{-3}h^{lm}\bigl[{4\over 3}\delta^b_a
(\Omega^{-1})_{;c(m}{}R^c_{~l}+\bigl((\Omega^{-1})_{;m}^{~~b}{}R_{al}
+ (\Omega^{-1})_{;al}{}R^b_{~m}\bigr)\nonumber\\
& +& {1\over
3}(\Omega^{-1})_{;mc}\bigl({R^b_{~a}}^c_{~l} +
R^{bc}_{~~al}\bigr)\bigr]\nonumber\\
&+& {\alpha\over
36}\Omega^{-4}(\Omega^2)_{;rs}{}g^{rs}\bigl[h^{lb} \Omega_{al} +
h^{lg}\bigl(g_{la} \Omega^b_{~g}-\delta^b_a \Omega_{gl}\nonumber\\
&-& g_{lg} \Omega^b_{~a}\bigr)\bigr]- {8\alpha\over
9}\Omega^{-5}h^{lm}\bigl[{3\over
4}\bigl((\Omega^{-1})_{;al}{}\delta^b_m + \delta^b_a
(\Omega^{-1})_{;ml}\nonumber\\
&+& g_{al}(\Omega^{-1})_{;m}^{~~b}\bigr)\bigr](\Omega^2)_{;rs}{}g^{rs}-{4\alpha\over
9} \Omega^{-6}
h^{lm}(\Omega^2)_{;rs}{}g^{rs}\biggl(\delta_a^b{}R_{lm}\nonumber\\
&+& {3\over 4}\bigl[R_{al}\delta^b_m + g_{al}{}R^b_{~m}\bigr] +
{1\over 4}R^b_{~mal}\biggr)\nonumber\\
&+&{\alpha\over 9}\Omega^{-8}
h^{lm}(\Omega^2)_{;rs}{}g^{rs}(\Omega^2)_{;tp}{}g^{tp}\bigl(\delta^b_a{}
g_{lm}+{3\over 2}\delta_m^b{}g_{al}\bigr)~,
\end{eqnarray}
where
\begin{equation}
\Omega^a_b = 4\Omega^{-1}(\Omega^{-1})_{;bc}{}g^{ae}
-2(\Omega^{-1})_{;c}(\Omega^{-1})_{;d}{}g^{cd}\delta^a_b~.
\end{equation}
Bearing in mind (\ref{weyl_trafo}), we have for a pure gravitational
field that
\bea
\tilde{C}_{ilmb} = \tilde{g}_{it} \tilde{C}^t_{~lmb} = \Omega^2
C_{ilmb}~, \\
\tilde{C}^{klm}_{~~~~a} = \Omega^{-4} C^{klm}_{~~~~a}~.
\eea
and so
\bea
\tilde{B}^k_{~iab} &=& \tilde{C}^{klm}_{~~~a} \tilde{C}_{ilmb} +
\tilde{C}^{klm}_{~~~b} \tilde{C}_{ilma} - \frac{1}{2} \delta_i^k
\tilde{C}^{lmn}_{~~~a} \tilde{C}_{lmnb} \nonumber \\
&=& \Omega^{-2} \left( C^{klm}_{~~~a} C_{ilmb} +
C^{klm}_{~~~b} C_{ilma} - \frac{1}{2} \delta_i^k
C^{lmn}_{~~~a} C_{lmnb} \right)~,
\eea
which means that the four times covariant form of the
Bel-Robinson tensor (\ref{bel}) for the pure gravitational field is
an invariant of the conformal transformation, i.e.,
\be
\tilde{B}_{kiab} = B_{kiab}~,
\ee
Similarly, for a pure gravity, from (\ref{pel}) and from the formulas given in
Section \ref{conrel}, one can easily obtain that
\begin{equation}
\label{pgSab}
_g\tilde{S}_a^{~b}(P;\tilde{v}^l) = \Omega^{-D}{} _g
S_a^{~b}(P;v^l),
\end{equation}
from which we have that
\begin{equation}
\sqrt{-\tilde{g}} _g \tilde{S}_a^{~b}(P;\tilde{v}^l) =
\sqrt{-g} _g S_a^{~b}(P;v^l),
\end{equation}
i.e., the tensorial density $\sqrt{-g}_g S_a^{~b}(P;v^l)$
is an invariant of the conformal transformation (\ref{conf_trafo}).
For a conformally flat spacetime and for a pure gravitational field we
have that
\begin{equation}
_g \tilde{S}_a^{~b} = \sqrt{-\tilde{g}} _g\tilde{S}_a^{~b}
= _g S_a^{~b} = \sqrt{-g} _g S_a^{~b} = 0~,
\end{equation}
which is, for example, the case of the Friedman universes.

\section{Angular Supermomentum tensors in Conformal Frames}
\label{sect6}

In this Section we further extend the notion of superenergy onto the
angular momentum which has been introduced in Ref. \cite{supermomentum}.

The canonical angular supermomentum tensors can be defined
in analogy to the canonical superenergy tensors as
\begin{equation}
S^{(a)(b)(c)}(P) = S^{abc}(P) :=\displaystyle\lim_{\Omega\to
P}{\int\limits_{\Omega}\bigl[M^{(a)(b)(c)}(y) -
M^{(a)(b)(c)}(P)\bigr]d\Omega\over
1/2\int\limits_{\Omega}\sigma(P;y)d\Omega},
\label{defSabc}
\end{equation}
where
\begin{equation}
M^{(a)(b)(c)}(y) := M^{ikl}(y) e^{(a)}_i(y) e^{(b)}_k(y) e^{(c)}_l(y),
\end{equation}
\begin{eqnarray}
M^{(a)(b)(c)}(P) &:=&  M^{ikl}(P) e^{(a)}_i(P) e^{(b)}_k(P) e^{(c)}_l(P) =
M^{ikl}(P)\delta^a_i\delta^b_k\delta^c_l \nonumber \\
&=& M^{abc}(P)
\label{moment}
\end{eqnarray}
are the physical (or tetrad) components of the field $M^{ikl}(y) =
- M^{kil}(y)$ which describe angular momentum densities. As in
(\ref{Sabtet})-(\ref{tetrad}),
$e^i_{(a)}(y), ~e^{(b)}_k(y)$ denote orthonormal bases such that
$e^i_{(a)}(P) = \delta^i_a$ and its dual are parallel propagated
along geodesics through {\bf P} and $\Omega$ is a sufficiently small
ball with centre at {\bf P}. As in Section \ref{sect5} we apply
the fact that at {\bf P} the tetrad and normal
components of an object are equal and so we again omit tetrad brackets for indices of
any quantity attached to the point {\bf P}.

For matter as $M^{ikl}(y)$ we take
\begin{equation}
_m M^{ikl}(y) = \sqrt{- g}\bigl(y^i T^{kl} - y^k T^{il}\bigr),
\label{defmom}
\end{equation}
where $T^{ik}$ are the components of a symmetric
energy-momentum tensor of matter and $y^i$ denote the normal
coordinates. The formula (\ref{defmom}) gives the total angular momentum densities,
orbital and spinorial because the dynamical energy-momentum tensor
of matter $T^{ik}$ comes from the canonical energy-momentum tensor by using
the Belinfante-Rosenfeld symmetrization procedure and, therefore,
includes the spin of matter \cite{BT}.
Note that the normal coordinates $y^i$ form the components of the
local radius-vector ${\vec y}$ with respect to the origin {\bf P}. In
consequence, the components of the $_m M^{ikl}(y)$ form a tensor
density.

For the gravitational field we take the gravitational angular
momentum pseudotensor of Bergmann and Thomson (\ref{BTdeffull}) to construct
\begin{equation}
_g M^{ikl}(y) = _F U^{i[kl]}(y) - _F U^{k[il]}(y) +\sqrt{\vert
g\vert}\bigl(y^i_{BT} t^{kl} - y^k_{BT} t^{il}\bigr),
\label{bergmom}
\end{equation}
where $_F U^{i[kl]} := g^{im} _FU_m^{~[kl]}$
are von Freud superpotentials (\ref{freud}).

In fact, the Bergmann-Thomson pseudotensor can be interpreted as the sum of the spinorial
part
\begin{equation}
S^{ikl} := _F U^{i[kl]} - _F U^{k[il]}
\end{equation}
and the orbital part
\begin{equation}
O^{ikl} := \sqrt{- g}\bigl(y^i _{BT} t^{kl} - y^k _{BT}
t^{il}\bigr)
\end{equation}
of the gravitational angular momentum densities.

Substitution of (\ref{defmom}) and (\ref{bergmom}) (expanded up to third
order) into (\ref{defSabc}) gives the canonical angular supermomentum tensors for
matter and gravitation, respectively \cite{supermomentum},
\begin{eqnarray}
\label{Sabcmatter}
_m S^{abc}(P;v^l)& = & 2\bigl[{h}^{ap} \nabla_p {} {T}^{bc}
 - {h}^{bp} \nabla_p {} {T}^{ac}\bigr],
\end{eqnarray}
\begin{eqnarray}
\label{Sabcgrav}
_g S^{abc}(P;v^l) & = &\alpha {h}^{pt} \bigl[\bigl({g}^{ac} {g}^{br} - {g}^{bc} {g}^{ar}\bigr){}\nabla_{(t} {R}_{pr)} \nonumber
\\
 & + & 2{g}^{ar}\nabla_{(t}{{R}^{(b}_{~~p}}{}^{c)}_{~~r)} - 2{g}^{br} \nabla_{(t} {{R}^{(a}_{~~p}}{}^{c)}_{~~r)}\nonumber \\
& + & {2\over 3} {g}^{bc}\bigl(\nabla_r{{R}^r_{~(t}}{}^a_{~p)} -
 \nabla_{(p} {R}^a_{t)}\bigr) \nonumber \\
& - & {2\over 3}{g}^{ac}\bigl(\nabla_r {{R}^r_{~(t}}{}^b_{~p)} -
\nabla_{(p} {R}^b_{t)}\bigr)\bigr].
\end{eqnarray}
Both these tensors are antisymmetric in the first two indices $S^{abc} = - S^{bac}$.
In vacuum, the gravitational canonical angular supermomentum tensor (\ref{Sabcgrav})
simplifies to
\begin{equation}
_g S^{abc}(P;v^l) = 2\alpha h^{pt} \bigl[g^{ar} \nabla_{(p}{R}^{(b}_{~~t}{}^{c)}_{~~r)}
- {g}^{br}\nabla_{(p}{{R}^{(a}_{~~t}}{}^{c)}_{~~r)}\bigr].
\end{equation}
Note that the orbital part $O^{ikl} = \sqrt{- g}\bigl(y^i _{BT}
t^{kl} - y^k _{BT} t^{il}\bigr)$ gives no contribution to $_g S^{abc}(P;v^l)$.
Only the spinorial part $S^{ikl} = _F
U^{i[kl]} - _F U^{k[il]}$ contributes. Also, the canonical angular supermomentum tensor $_g S^{abc}(P;v^l)$
and $_m S^{abc}(P;v^l)$ of gravitation and matter do not require the introduction
of the notion of a radius vector.

After some algebra, one may show that there are the following transformational rules
for the angular supermomentum tensors of  matter $_m S^{ikl}(P;v^a)$ and for pure
gravitation (which is composed of the Weyl tensor only)
$_g S^{ikl}(P;v^a)$ under conformal transformation (\ref{conf_trafo}):
\begin{eqnarray}
\label{mSikl}
_m \tilde{S}^{ikl}(P;\tilde{v}^a)&=& \Omega^{-8} {_m} S^{ikl}(P; v^a) +  4
\Omega^{-8}\bigl[h^{p[i}\bigl(P^{k]}_{~~pr}{}T^{rl} \nonumber\\
&+&P^l_{~pr}{}T^{k]r}\bigr)\bigr] -24\Omega^{-9}
h^{p[i}\Omega_{,p}{}T^{k]l};
\end{eqnarray}
\begin{eqnarray}
\label{gSikl}
_g S^{ikl}(P;\tilde{v}^a) &=& \Omega^{-6} {_g} S^{ikl}(P;v^a)+
2\alpha\Omega^{-6}h^{pt}\bigl\{g^{ib}\bigl[P^{(l}_{~~(t\vert
s\vert}{}C^{\vert s\vert}_{~~~b}{}^{k)}_{~~p)}\nonumber\\
&+& P^{(k}_{~~(t\vert s\vert}{}C^{l)}_{~~b}{}^s_{~p)}-
P^s_{~(tb}{}C^{(l}_{~\vert s\vert}{}^{k)}_{~~p)} -
P^s_{~(tp}{}C^{(l}_{~~b)}{}^{k)}_{~~s}\bigr]\nonumber\\
&-& g^{kb}\bigl[P^{(l}_{~~(t\vert s\vert}{} C^{\vert
s\vert}_{~~~b}{}^{i)}_{~~p)}+ P^{(i}_{~~(t\vert s\vert}{}
c^{l)}_{~~b}{}^s_{~p)} - P^s_{~(tb}{} C^{(l}_{~~\vert
s\vert}{}^{i)}_{~~p)}\nonumber\\
&-& P^s_{~(tp}{}
C^{(l}_{~~b)}{}^{i)}_{~~s}\bigr]\bigr\}\nonumber\\
&-&4\alpha\Omega^{-7}
h^{tp}\bigl[g^{ib}\Omega_{,t}{}C^{(l}_{~b}{}^{k)}_{~~p)}
-g^{kb}\Omega_{,t}{}C^{(l}_{~~b}{}^{i)}_{~~p)}\bigr]~,
\end{eqnarray}
with $h^{lm}$ and $P^a_{~bc}$ given by (\ref{hlm}) and
(\ref{Pabc}) and $C^a_{~bcd}$ being the components of the Weyl conformal curvature
tensor (\ref{weyl_def}).
It is obvious that in a conformally flat spacetime, one has
\begin{equation}
_g \tilde{S}^{ikl} = _g S^{ikl} =0.
\end{equation}

\section{Conclusion}

In this paper we have analyzed the rules of the conformal transformations
of the energetic and superenergetic quantities which were proposed in general relativity
and can be applied to some extended theories of gravity in which
physics is studied in different conformal frames.

In particular, we have found the rules of the conformal transformation
for the energetic quantities such as the Einstein energy-momentum
complex, the Bergmann-Thomson angular momentum complex, the
superenergy tensor, and the angular supermomentum tensor of gravitation and matter.

We have shown that the conformal transformation rules for the matter
parts of both the Einstein complex and the Bergmann-Thomson complex are
fairly simple (Eqs. (\ref{mTik}) and (\ref{mMijk})), while the transformation rules for their
gravitational parts are more complicated (Eqs. (\ref{pseudoE}) and (\ref{gMijk})).
We have also found that the transformational rules of
the superenergy tensor of matter (\ref{mSab}) and the superenergy tensor of gravity
(\ref{gSab}) are quite complicated, except for the case of a pure gravity (\ref{pgSab}). In such a special
case the superenergy density as well as the sum of the matter energy density and the
superenergy density are invariants of the conformal transformation, i.e.
\bea
\sqrt{-\tilde{g}} \left( \tilde{T}_i^k + _g \tilde{S}_i^k \right)
= \sqrt{-g} \left( T_i^k~ + _g S_i^k \right). \nonumber
\eea
Besides, in that case (of a pure gravity), a conformal invariant is also the
Bel-Robinson tensor
\bea
\tilde{B}_{kiab} = B_{kiab}~, \nonumber
\eea
which is a part of the superenergy tensor. As for the angular supermomentum
tensor of gravity - it emerges that
its transformational rule (\ref{gSikl}), even for a pure gravity, is quite
complicated. This, however, is not the case for the angular supermomentum tensor of
matter (Eq. (\ref{mSikl})).

Some other remarks from our investigations are as follows. The
conformal transformation rule of the Einstein pseudotensor
(\ref{pseudoE}) vastly simplifies in $D=2$ dimensional spacetime.
This seem to reflect the fact that all the two-dimensional
manifolds are conformally flat. On the other hand, both a pure gravity
superenergy tensor and a pure gravity angular supermomentum tensor
vanish in all conformally flat spacetimes.

Because the superenergetic quantities are constructed of some
combinations of the geometric quantities, we have studied the rules
of their transformations from one conformal frame to another. In
particular, we have derived the rules of the conformal
transformation for the curvature invariants $R^2$, $R_{ab}R^{ab}$,
$R_{abcd} R^{abcd}$ (Eqs. (\ref{R2})-(\ref{Riem2})) and the Gauss-Bonnet invariant
$R_{GB} = R_{abcd} R^{abcd} - 4 R_{ab} R^{ab} + R^2$ (Eqs. (\ref{RGBtil})-(\ref{RGB}))
in an arbitrary spacetime dimension.

All the rules we found would be applied to the discussion of the
conformal transformation of energetic and superenergetic
quantities in some special models of spacetime \cite{prepar}.
Especially, these rules should be very effective to calculate
energetic and superenergetic quantities in conformally flat
spacetimes.

Quite recently, the form of the energy-momentum complexes within
the framework of extended gravity $f(R)$ theories have been studied \cite{vagenas}.
In fact, it is very common to use conformal frames (Jordan and Einstein) in presentation
of these theories \cite{f(R)} so that our results of Sections \ref{sect3} and \ref{sect4}
can be applied to study the sensitivity
of the energy-momentum complexes to conformal transformations in these more general theories
of gravity. On the other hand, since the complexes are sensitive to coordinate
transformations, then it would be much better to study the superenergy and the
angular supermomentum (which are covariant) in $f(R)$ theories of gravity with the stress
onto the problem of their sensitivity to conformal transformations.

\section{Acknowledgments}

The authors acknowledge partial support of the Polish Ministry of
Education and Science grant No 1 P03B 043 29 (years 2005-2007).
We thank Adam Balcerzak for assistance in deriving one of the formulas.


\end{document}